\documentclass[
superscriptaddress,aps,preprintnumbers,amsmath,showpacs,amssymb,prl,reprint]{revtex4-1}

\usepackage[dvipdfm]{graphicx}
\usepackage{amsfonts}
\usepackage[mathscr]{eucal}
\usepackage{ascmac}
\usepackage{epstopdf}
\usepackage{dcolumn}
\usepackage{bm}
\usepackage{hyperref}
\usepackage{color}

\begin{document}


\def\a{\alpha}
\def\b{\beta}
\def\c{\varepsilon}
\def\d{\delta}
\def\e{\epsilon}
\def\f{\phi}
\def\g{\gamma}
\def\h{\theta}
\def\k{\kappa}
\def\l{\lambda}
\def\m{\mu}
\def\n{\nu}
\def\p{\psi}
\def\q{\partial}
\def\r{\rho}
\def\s{\sigma}
\def\t{\tau}
\def\u{\upsilon}
\def\v{\varphi}
\def\w{\omega}
\def\x{\xi}
\def\y{\eta}
\def\z{\zeta}
\def\D{\Delta}
\def\G{\Gamma}
\def\H{\Theta}
\def\L{\Lambda}
\def\F{\Phi}
\def\P{\Psi}
\def\S{\Sigma}

\def\o{\over}
\def\beq{\begin{eqnarray}}
\def\eeq{\end{eqnarray}}
\newcommand{\gsim}{ \mathop{}_{\textstyle \sim}^{\textstyle >} }
\newcommand{\lsim}{ \mathop{}_{\textstyle \sim}^{\textstyle <} }
\newcommand{\vev}[1]{ \left\langle {#1} \right\rangle }
\newcommand{\bra}[1]{ \langle {#1} | }
\newcommand{\ket}[1]{ | {#1} \rangle }
\newcommand{\EV}{ \ {\rm eV} }
\newcommand{\KEV}{ \ {\rm keV} }
\newcommand{\MEV}{\  {\rm MeV} }
\newcommand{\GEV}{\  {\rm GeV} }
\newcommand{\TEV}{\  {\rm TeV} }
\newcommand{\1}{\mbox{1}\hspace{-0.25em}\mbox{l}}
\def\diag{\mathop{\rm diag}\nolimits}
\def\Spin{\mathop{\rm Spin}}
\def\SO{\mathop{\rm SO}}
\def\O{\mathop{\rm O}}
\def\SU{\mathop{\rm SU}}
\def\U{\mathop{\rm U}}
\def\Sp{\mathop{\rm Sp}}
\def\SL{\mathop{\rm SL}}
\def\tr{\mathop{\rm tr}}

\def\IJMP{Int.~J.~Mod.~Phys. }
\def\MPL{Mod.~Phys.~Lett. }
\def\NP{Nucl.~Phys. }
\def\PL{Phys.~Lett. }
\def\PR{Phys.~Rev. }
\def\PRL{Phys.~Rev.~Lett. }
\def\PTP{Prog.~Theor.~Phys. }
\def\ZP{Z.~Phys. }

\def\dd{\mathrm{d}}
\def\ff{\mathrm{f}}
\def\BH{{\rm BH}}
\def\inf{{\rm inf}}
\def\ev{{\rm evap}}
\def\eq{{\rm eq}}
\def\SM{{\rm sm}}
\def\Mpl{M_{\rm Pl}}
\def\GeV{ \ {\rm GeV}}
\newcommand{\Red}[1]{\textcolor{red}{#1}}

\def\mDM{m_{\rm DM}}
\def\mphi{m_{\text{I}}}
\def\TeV{\ {\rm TeV}}
\def\MeV{\ {\rm MeV}}
\def\Gphi{\Gamma_\phi}
\def\TR{T_{\rm RH}}
\def\Br{{\rm Br}}
\def\DM{{\rm DM}}
\def\Eth{E_{\rm th}}
\newcommand{\lmk}{\left(}  
\newcommand{\rmk}{\right)}
\newcommand{\lkk}{\left[}  
\newcommand{\rkk}{\right]}
\newcommand{\lhk}{\left \{ }  
\newcommand{\rhk}{\right \} }
\newcommand{\del}{\partial}  
\newcommand{\la}{\left\langle} 
\newcommand{\ra}{\right\rangle}

\newcommand{\qel}{\hat{q}_{el}}
\newcommand{\ksplit}{k_{\text{split}}}
\def\GDM{\Gamma_{\text{DM}}}
\newcommand{\half}{\frac{1}{2}}
\def\Gsplit{\Gamma_{\text{split}}}

\def\mg{m_{3/2}}
\newcommand{\abs}[1]{\left\vert {#1} \right\vert}
\def\Im{{\rm Im}}
\def\bea{\begin{array}}
\def\eea{\end{array}}
\def\Mpl{M_{\text{pl}}}
\def\mN{m_{\text{NLSP}}}
\def\Td{T_{\text{decay}}}
\def\mphi{m_{\phi}}
\def\tanb{\text{tan}\beta}
\def\signmu{\text{sign}[\mu]}
\def\fb{\text{ fb}}
\def\osc{\text{osc}}


\title{
A solution to the baryon-DM coincidence problem 
in~the~CMSSM 
with~a~126-GeV~Higgs~boson
}

\author{Ayuki Kamada}
\affiliation{Kavli IPMU (WPI), TODIAS, University of Tokyo, Kashiwa, 277-8583, Japan}
\affiliation{Department of Physics and Astronomy, University of California, Riverside, CA, 92507, USA}
\author{Masahiro Kawasaki}
\affiliation{Kavli IPMU (WPI), TODIAS, University of Tokyo, Kashiwa, 277-8583, Japan}
\affiliation{ICRR, University of Tokyo, Kashiwa, 277-8582, Japan}
\author{Masaki Yamada}
\affiliation{Kavli IPMU (WPI), TODIAS, University of Tokyo, Kashiwa, 277-8583, Japan}
\affiliation{ICRR, University of Tokyo, Kashiwa, 277-8582, Japan}
\begin{abstract}
We show that the baryon-dark matter coincidence problem is solved 
in the CMSSM. 
The baryons and dark matter 
are generated simultaneously through the late-time decay of non-topological solitons, Q-balls, 
which are formed after the Affleck-Dine baryogenesis. 
A certain relation between the universal scalar mass, $m_0$, and the universal gaugino mass, $M_{1/2}$, 
is required to solve the coincidence problem, marginally depending on the other CMSSM parameters, 
and the result is consistent with the observation of the 126-GeV Higgs boson.

\end{abstract}

\date{\today}
\pacs{98.80.Cq, 95.35.+d, 12.60.Jv}
\maketitle
\preprint{IPMU 14-0127}
\preprint{ICRR-Report-683-2014-9}

\noindent {\bf Introduction. }
The origins of the baryon asymmetry and dark matter (DM) are 
mysteries with which particle physics and cosmology are confronted. 
The precision measurement of the fluctuation of the cosmic background radiation 
shows a coincidence between the relic density of baryon and DM: 
$\Omega_{\DM}/ \Omega_b \simeq 5$~\cite{Hinshaw:2012aka, Ade:2013zuv}, 
referred to as the baryon-DM coincidence problem. 
This coincidence implies that baryon and DM are generated from the same origin. 
These observations require new physics beyond the Standard Model 
and 
give us useful information to close in on the true model of particle physics.

Low-energy supersymmetric (SUSY) models 
are well-motivated in particle physics 
in light of 
a gauge coupling unification and 
a solution of hierarchy problem between the electroweak scale and the Planck scale. 
The discovery of the 126-GeV Higgs boson by the LHC experiment~\cite{Aad:2012tfa, Chatrchyan:2012ufa}
and theoretical 3-loop calculations of Higgs mass implies that 
the masses of SUSY particles are $O(1)$ TeV~\cite{Feng:2013tvd, Buchmueller:2013psa}. 
In SUSY theories, 
the lightest SUSY particle (LSP) is a good candidate for DM, 
and the baryon asymmetry can be generated by the Affleck-Dine baryogenesis~\cite{AD, DRT}.

Based on a variant of the Affleck-Dine baryogenesis, 
a scenario for co-genesis of baryon and DM has been proposed by 
Enqvist and McDonald
to overcome the baryon-DM coincidence problem 
in models of gravity mediation~\cite{EnMc}. 
The baryon asymmetry is generated as a form of squark condensation, 
which then fragments into long-lived non-topological solitons, 
referred to as Q-balls~\cite{Coleman, Qsusy, KuSh, KK1, KK2, KK3}. 
Eventually, each Q-ball releases its baryon charge from its surface (evaporation) 
through baryon-number-conserving elementary processes, 
such as the decay of squark into quark and gaugino~\cite{evap}. 
Since the gaugino soon decays into the LSP DM, 
the baryon (quarks) and DM (LSPs) are generated simultaneously by the evapolation of Q-balls. 
Enqvist and McDonald focused on the process of squark decay into quark and gaugino, 
which implies that the number of quarks is the same as that of the LSP due to the R-parity conservation. 
To explain the baryon-DM coincidence problem, 
they concluded that the mass of DM has to be $O(1) \GEV$~\cite{EnMc}. 
Since this is excluded by the collider experiments, 
alternative scenarios have been proposed in Refs.~\cite{Fujii:2001xp, Fujii:2002kr, 
Roszkowski:2006kw, ShKu, Kasuya:2011ix, Doddato:2012ja, Kasuya:2012mh}.

Recently, we have investigated the evapolation of Q-ball in detail 
and have found that the process of squark annihilation $\tilde{q} \tilde{q} \to q q$ 
is the dominant process for 
the evapolation of Q-balls~\cite{KY}. 
This implies that the DM mass of $O(100)$ GeV is consistent with 
the observed DM abundance 
even when the annihilation of DM is ineffective~\cite{KKY}. 
In that work, we have considered a model of wino-LSP 
so that its thermal relic density can be neglected. 
In this letter, 
we apply the calculation to the CMSSM with a low reheating temperature, 
which dilutes the thermal relic density of bino-LSP. 
Our results clearly shows that 
the baryon-DM coincidence problem can be solved 
even in such a simple model without any additional fields and without any fine-tuned parameters. 
The CMSSM parameter space for co-genesis is unique from and complementary to the conventional bino
thermal relic scenarios. 
In particular, co-genesis predicts lighter sfermions and heavier gauginos than conventional scenarios. 
The results are consistent with the observation of the 126-GeV Higgs boson 
and would be tested by future LHC experiments.

\vspace{0.2cm}
\noindent {\bf Origin of baryon asymmetry. }
Let us consider the dynamics of a F- and D-flat direction denoted by $\phi$ 
carrying nonzero baryon charge $b$. 
During inflation, the flat direction obtains Hubble-induced terms 
through the supergravity effect, 
and its potential is given as 
\beq
 V &=& m_\phi (\phi)^2 | \phi |^2 
 + c_H H^2 \abs{\phi}^2 
 + \frac{\lambda^2}{\Mpl^{2n-6}} \abs{\phi}^{2(n-1)} \nonumber \\
&+& \lmk \frac{- \lambda a_g}{n \Mpl^{n-3}} \mg \phi^n 
+ \frac{ - \lambda a_H}{n \Mpl^{n-3}} H \phi^n 
+ \text{h.c.} \rmk, 
\eeq
where $H$ is the Hubble parameter, $\Mpl$ ($\simeq 2.4 \times 10^{18} \GEV$) is the Planck scale, 
and 
the parameters $c_H$, $a_g$, and $a_H$ 
are $O(1)$ constants. 
Here we have included higher dimensional terms coming from a superpotential 
$W = \lambda \phi^n / n \Mpl^{n-3}$. 
We write the soft mass of the flat direction as $\mphi (\phi)$, 
implicitly taking into account the logarithmic dependence on $\phi$ due to quantum correction. 
The gravitino mass $\mg$ is the same order as $\mphi$ in models of gravity mediation.

Hereafter, we assume $c_H < 0$, 
which makes the flat direction obtain a large vacuum expectation value (VEV) 
determined by the unknown parameters $\lambda$ and $n$. 
The phase direction stays at a certain phase 
determined by Im[$a_H$]. 
After inflation ends, the Hubble parameter decreases with time as $\propto a^{-3/2}$, 
where $a$ is a scale factor. 
Once the Hubble parameter $H$ decreases down to $\mphi$,
the flat direction feels the soft mass term and starts to oscillate around the low energy vacuum. 
At the same time, 
the phase direction feels the soft A-term and 
is kicked into phase direction 
due to Im[$a_H$] $\ne$ Im[$a_g$]. 
This rotation in the complex plane results in the generation of baryon asymmetry, 
which is given as 
\beq
\lmk \frac{a(t)}{a(t_{\osc})} \rmk^3 n_B (t) 
&=& 2 b \int a^3 \Im \lkk \phi^* \frac{\del V}{\del \phi^*} \rkk \dd t \nonumber \\
&\sim&
b \mg \abs{\phi_{\osc}}^2. 
\label{n_b}
\eeq
Without any entropy production other than the reheating of the Universe due to inflaton decay, 
the baryon-to-entropy ratio is given as 
\beq
Y_B \equiv \frac{n_B}{s} 
= \frac{3T_{\rm RH}}{4} \left.\frac{n_B}{\rho_{\rm inf}}\right|_{\rm osc}
\sim  \frac{b \mg T_{\rm RH}}{4 \mphi^2} \lmk \frac{ |\phi_{\osc}|}{ \Mpl} \rmk^2, 
\label{Y_B}
\eeq
where $\rho_{\rm inf}$ is the energy density of inflaton. 
Note that the uncertainty in the last equality can be 
compensated by the choice of the reheating temperature $\TR$, 
to which our final predictions are insensitive as shown below.

Since 
the usual thermal relic density of the bino-LSP is over-abundant 
without help of slepton/stop co-annihilation or large higgsino mixing, 
co-genesis can be realized only with 
a reheating temperature 
lower than the LSP freeze-out temperature 
to dilute the thermal relic density of the LSP. 
Note that the abundance of DM produced through the reheating process is negligible 
for a sufficiently low reheating temperature $\TR \lesssim 100 \MeV$ and 
a sufficiently large inflaton mass~\cite{HKMY}, 
though these condition have 
$O(1)$ uncertainty. 
Such a low reheating temperature is also favoured 
in light of baryonic isocurvature constraints~\cite{Enqvist:1998pf, Enqvist:1999hv, Kawasaki:2001in, Kasuya:2008xp, Harigaya:2014tla}. 
Hereafter, we take the reheating temperature as a free parameter less than $O(1)$ GeV. 
Once we determine the reheating temperature, 
the observed baryon-to-entropy ratio ($Y_B = 8.7 \times 10^{-11}$~\cite{pdg}) determines the VEV of the flat direction $\phi_{\osc}$ 
through Eq.~(\ref{Y_B}).

\vspace{0.2cm}
\noindent {\bf Q-ball. }
The amplitude of the flat direction soon decreases due to the Hubble expansion 
after starting the oscillation. 
The dynamics of the flat direction is then determined by the soft mass term as 
\begin{equation}
 V = m_\phi^2 (\phi) \abs{\phi}^2 
 = m^2_\phi \vert \phi \vert^2 \left( 1+ K \log \frac{ \vert \phi \vert^2}{\Mpl^2} \right), 
 \label{Q-ball potential}
\end{equation}
where we explicitly write the running feature of the mass of the flat direction. 
In most cases we are interested in, 
the gauge interaction dominantly induces a negative $K$, 
as shown below. 
The negative $K$ leads to an instability of 
the homogeneous solution of the squark condensation 
and results in the formation of non-topological solitons called Q-balls, 
which are localized lumps of squark condensation 
carrying large baryon charges~\cite{EnMc, Coleman, Qsusy, KuSh, KK1, KK2, KK3}.

The first stage of Q-ball formation can be investigated by the linear analysis~\cite{EnMc}. 
Small fluctuations over a homogeneous background are unstable and grows exponentially 
at wave number $k < k_{\rm max} \simeq 2 \abs{K}^{1/2} \mphi$. 
In particular, the mode of $\abs{K}^{1/2} \mphi$ grows most efficiently and 
Q-balls with a size of $R \simeq 1/\abs{K}^{1/2} \mphi$ 
are formed. 
A typical charge ($=$ baryon number) of Q-balls is 
thus estimated as $Q \sim R^3 n_B (t_{\rm form})$, 
where $t_{\rm form}$ is the time of Q-ball formation. 
In fact, numerical simulations of Q-ball formation imply 
that it is given as 
\beq
 Q  =
  \beta b
 \lmk \frac{\phi_{\osc}}{\mphi} \rmk^2, 
 \label{Q charge}
\eeq
where $\beta \simeq 0.02$~\cite{KK2, Hiramatsu:2010dx}.

The configuration and properties of Q-ball are obtained by deriving the configuration 
which minimizes the energy with a finite baryon charge. 
In the case of potential of Eq.~(\ref{Q-ball potential}), 
the configuration of the Q-ball is approximately gaussian~\cite{EnMc}: 
\beq
 \phi(r, t ) \simeq \frac{1}{\sqrt{2}} \phi_0 e^{- r^2 / 2 R^2} e^{-i \omega_0 t},
\eeq
where $R$, $\omega_0$, and $\phi_0$ are given as
\begin{eqnarray}
R &\simeq& \frac{1}{ \vert K \vert^{1/2} m_\phi (\phi_0)}, \label{gravproperty} \\
 \omega_0 &\simeq& m_\phi (\phi_0), \\
\phi_0 &\simeq& \lmk \frac{\vert K \vert}{\pi} \rmk^{3/4} m_\phi (\phi_0) Q^{1/2}. 
\end{eqnarray}
The energy of the Q-ball per unit charge is approximately equal to 
$\omega_0$.

\vspace{0.2cm}
\noindent {\bf Q-ball decay. }
Since Q-balls consist of squarks, 
each Q-ball gradually releases its baryon charge through the processes like $\tilde{q} \to q + \tilde{g}$ 
and $\tilde{q} \tilde{q} \to q q$, 
where $\tilde{q}$, $q$, and $\tilde{g}$ represent a squark, quark, 
and gaugino (or higgsino), respectively~\cite{evap, KY}. 
These reaction rates are so large that 
quark, gaugino, and higgsino production rates from Q-ball decay 
are saturated by the upper bound due to the Pauli blocking effect at the Q-ball surface~\cite{evap, KY}.

Since particles interacting with a Q-ball in tree level 
obtain the effective mass of $g_i \phi (r)$ inside the Q-ball, 
where $g_i$ generically represents a coupling constant, 
they are produced at the surface of the Q-ball 
with the effective radius $\tilde{R}_i$ determined by the condition $g_i \phi (\tilde{R}) = \omega_0$. 
This implies that the Q-ball decay rate is given as 
\beq
 \frac{\dd Q}{\dd t} \simeq \sum_i b_i 4 \pi \tilde{R}_i^2 \bm{n \cdot j_i}, \label{dNdt}
\eeq
where $\bm{n}$ is a pointing normal for the Q-ball surface 
and $\bm{j_i}$ is flux of each particle (baryon number $b_i$) interacting with the Q-ball. 
Q-balls completely evaporate at the time of $H \simeq Q^{-1} \dd Q/ \dd t$.

Let us consider the flux of quarks from Q-ball decay. 
They are mainly produced through the squark annihilation $\tilde{q} \tilde{q} \to q q$ 
and their flux determined by the Pauli blocking effect 
is given as 
\beq
 \bm{n \cdot j_i} &\simeq& 2 \int \frac{\dd^3 k }{(2 \pi)^3} \theta \lmk \omega_0 - 
 \abs{\bm{k}} \rmk \theta \lmk \bm{k \cdot n} \rmk \bm{\hat{k} \cdot n} \nonumber \\
 &=& \frac{\omega_0}{12 \pi^2}. 
 \label{flux}
\eeq
We can estimate the number of quarks ($n_q$) interacting with the Q-ball 
to take the summation in Eq.~(\ref{dNdt}) for a given flat direction. 
In general, 
there is an upper bound on $n_q$: 
$3 \text{ (color)} \times 6 \text{ (family)} \times 2 \text{ (chirality)} = 36$. 
The estimation of $n_q$ allows us to approximate 
the Q-ball decay rate such as 
$n_q/3 \times 4 \pi \tilde{R} \omega_0 /12\pi^2$, 
where $\tilde{R} \simeq (\log [ \phi_0/\omega_0 ] )^{1/2} R$.

Gauginos and/or higgsinos are produced from Q-ball 
through the squark decay ($\tilde{q} \to q + \tilde{g}$). 
In this case, we have to take into account 
the masses of SUSY particles. 
When a gaugino with a mass of $m_\chi$ obtains an energy $E_\chi$, which is within the interval of $[m_\chi, \omega_0]$, 
the relevant quark obtains the energy of $\omega_0 - E_\chi$. 
Their phase space integrals are then proportional to $\int \dd E_\chi p_i^2$, where $p_i$ 
represents the momentum of each particle. 
Since the flux is determined by the smaller phase space due to the Pauli exclusion principle, 
it is given as 
\beq
 \bm{n \cdot j_\chi} 
 &\simeq&
  \frac{1}{8 \pi^2}
  \int^{\omega_0}_{m_{\chi}} \text{d}E_\chi
    \min \lkk (\omega_0 - E_\chi)^2, E_\chi^2 - m_\chi^2 \rkk \label{flux2}\\
&=& \frac{\omega_0^3}{96 \pi^2} 
  \times f(m_{\chi}/\omega_0), \label{bino massive}\\[0.6em]
f(x ) &\equiv& 
	\begin{cases}
	  1- 6 x^2 + 8 x^3 - 3 x^4 &\text{for}~~ 0 \le  x \le 1, \\
	  0  &\text{for}~~ 1 <  x, 
	 \end{cases}
	  \label{correction}
\eeq
where we define a function $f(x)$ so that it can be applied to the case of $m_\chi \ge \omega_0$.

\vspace{0.2cm}
\noindent {\bf The scenario for co-genesis. }
Using Eqs.~(\ref{flux}) and (\ref{bino massive}), 
we obtain the baryon-to-DM ratio from Q-ball decay such as 
\beq
 \frac{\Omega_{\DM}}{\Omega_B} 
 &\simeq& 
 \frac{3 m_{\text{LSP}}}{m_p} 
 \frac{\Br(\text{Q-ball} \to \text{sparticles})}{\Br(\text{Q-ball} \to \text{quarks})} \nonumber \\
 &\simeq& 
 \frac{3 m_{\text{LSP}}}{m_p} 
  \frac{\sum_s g_s f (m_{\text{s}} / \omega_0)}{8 n_q},
 \label{bDMratio}
\eeq
where $m_\text{LSP}$ and $m_p$ are the LSP mass and the proton mass, respectively, 
and the summation is taken for gauginos and higgsinos which interact with the Q-ball. 
The number of degrees of freedom $g_s$ is 1 for the bino, 3 for the wino, 8 for the gluino, 
and 4 for the higgsino. 
We should emphasize that the resulting baryon-to-DM ratio Eq.~(\ref{bDMratio}) 
depends only on the masses of SUSY particles except for $n_q$, 
which is typically $O(10)$. 
Thus, we can calculate the baryon-to-DM ratio from Eq.~(\ref{bDMratio}) once we specify the SUSY model 
without any additional assumptions.

Here let us check whether the annihilation of the LSP is efficient or not, 
which might affect the baryon-to-DM ratio of Eq.~(\ref{bDMratio}). 
One might wonder that the spatial distribution of LSPs are localized 
around the Q-balls. 
However, 
the spatial distribution of LSPs becomes homogeneous 
due to their free streaming before 
its thermalization and annihilation become effective~\cite{EnMc, Doddato:2012ja}. 
The annihilation effect is ineffective when $n_{\rm DM} \la \sigma v \ra \lesssim H$ 
is satisfied at the time of Q-ball decay. 
In that case, the baryon-to-DM ratio given by Eq.~(\ref{bDMratio}) is justified. 
We estimate the annihilation cross section $\la \sigma v \ra$ as follows.

First, we check whether the LSPs 
kinematically thermalized 
due to elastic interactions 
with the thermal plasma or not. 
Since the LSP is mostly bino, 
the elastic scattering through sfermion exchange is in charge of losing their energy.  
The energy-loss rate is given as 
\beq
&& -\frac{\dd E_{\tilde b}}{E_{\tilde b} \dd t} 
= 
\sum_i
\frac{31\pi^{3}}{5120} g_{1}^{4} \frac{T^{6}}{m_{\tilde b}^{6}} E_{\tilde b}
\left(1 - \frac{m_{\tilde b}^{2}}{E_{\tilde b}^{2}} \right) \left ( 6\frac{E_{\tilde b}^{2}}{m_{\tilde b}^{2}} - 1 \right ) \notag \\
&& \qquad \times 
\left \{ \left( \frac{Y_{L}^{2}}{m_{{\tilde f}_{L}}^{2}/m_{\tilde b}^{2}-1} \right)^{2} + \left( \frac{Y_{R}^{2}}{m_{{\tilde f}_{R}}/m_{\tilde b}^{2}-1} \right)^{2} \right\} \,,
\eeq 
with $g_{1}=\sqrt{5/3}g_{Y}$, left/right-handed sfermion masses $m_{{\tilde f}_{L/R}}$, $Y_{L}=-1/2,1/6$ for leptons and quarks, and $Y_{R}=-1,2/3,-1/3$ for charged leptons, up- and down-type quarks.
The summation is taken for all relativistic particles. 
We average the energy-loss rate over non-thermal distribution, 
that is, we integrate it 
in terms of the energy of the bino $E_{\tilde{b}}$ 
with the weight given as the integrant of Eq.~(\ref{flux2}).

If the energy-loss is larger than the Hubble expansion rate, 
we use the thermally averaged annihilation cross section. 
Otherwise, we adopt the non-thermal annihilation cross section. 
The sfermion exchange dominates the annihilation of LSPs, 
whether the produced LSPs are thermalised or not. 
The annihilation cross sections are too lengthy to be presented here. 
For the thermally averaged annihilation cross section, 
we consider s- and p-wave contributions~\cite{Falk:1993we}. 
When we calculate the non-thermal annihilation cross section, 
we ignore the fermion masses and average it over non-thermal distribution.
We also take into account enhancement of annihilation cross section 
due to resonance effects 
though we find it sub-dominant.

\vspace{0.2cm}
\noindent {\bf Application to the CMSSM. }
Here we apply the above scenario in the CMSSM, 
where 
all MSSM parameters are determined by 
the universal scalar mass, $m_0$, 
the universal gaugino mass, $M_{1/2}$, 
the universal trilinear scalar coupling, $A_0$, 
the ratio of the VEV of the two Higgs fields, tan$\beta$, 
and the sign of the higgsino mass parameter, sign$[\mu]$. 
We use the code ${\tt SOFTSUSY~3.3.6}$ 
to calculate the spectrum of SUSY particles~\cite{Allanach:2001kg}. 
We estimate the parameters $K$ and $\mphi (\phi_0)$ 
as averages of beta functions and masses over all squarks 
at the energy scale of $\phi_0$, respectively.

\begin{figure}[!]
\centering 
\includegraphics[width=.45\textwidth, bb=0 0 360 356]{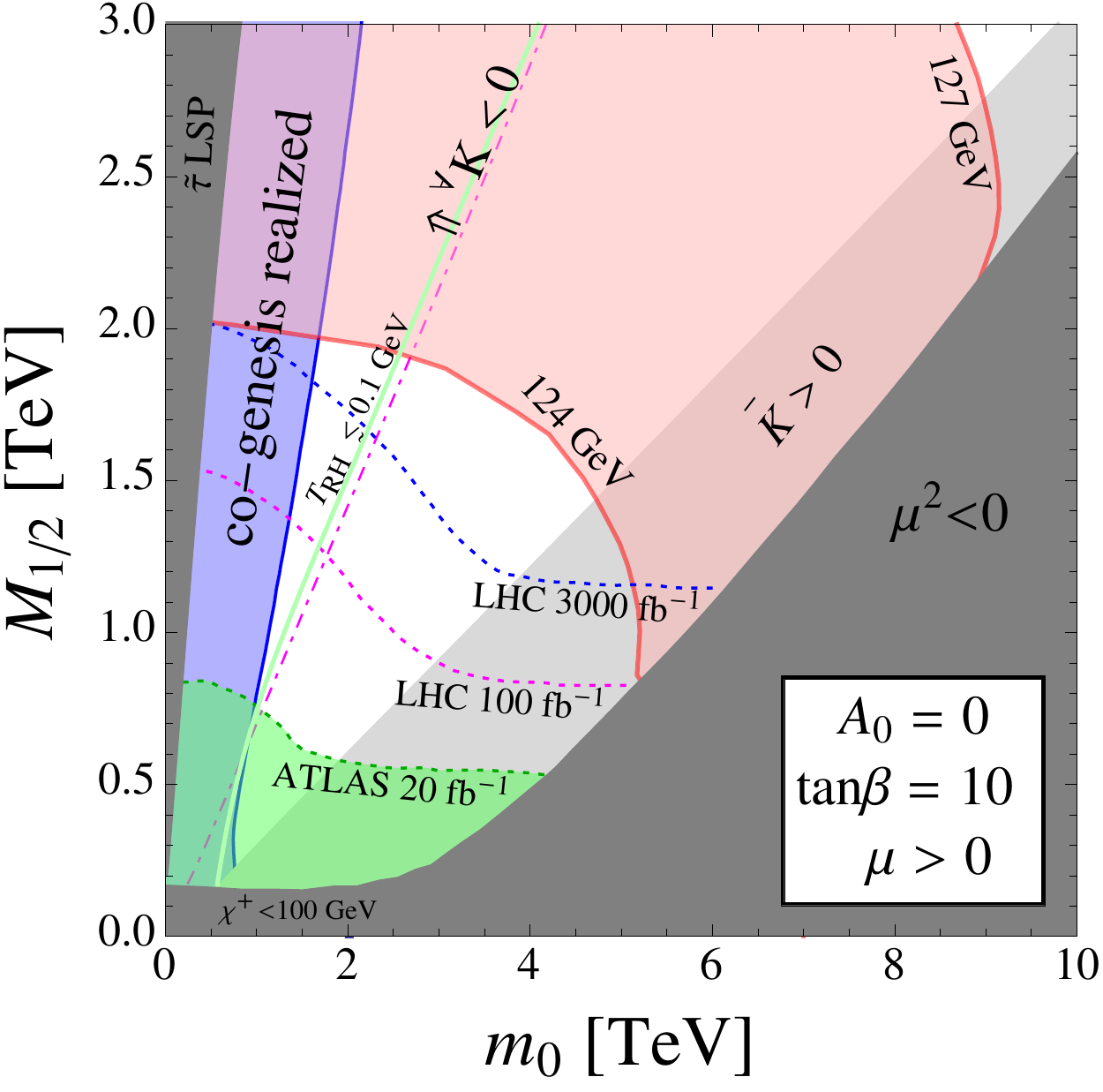} 
\vspace{0.5cm}
\hfill
\includegraphics[width=.45\textwidth, bb=0 0 360 354]{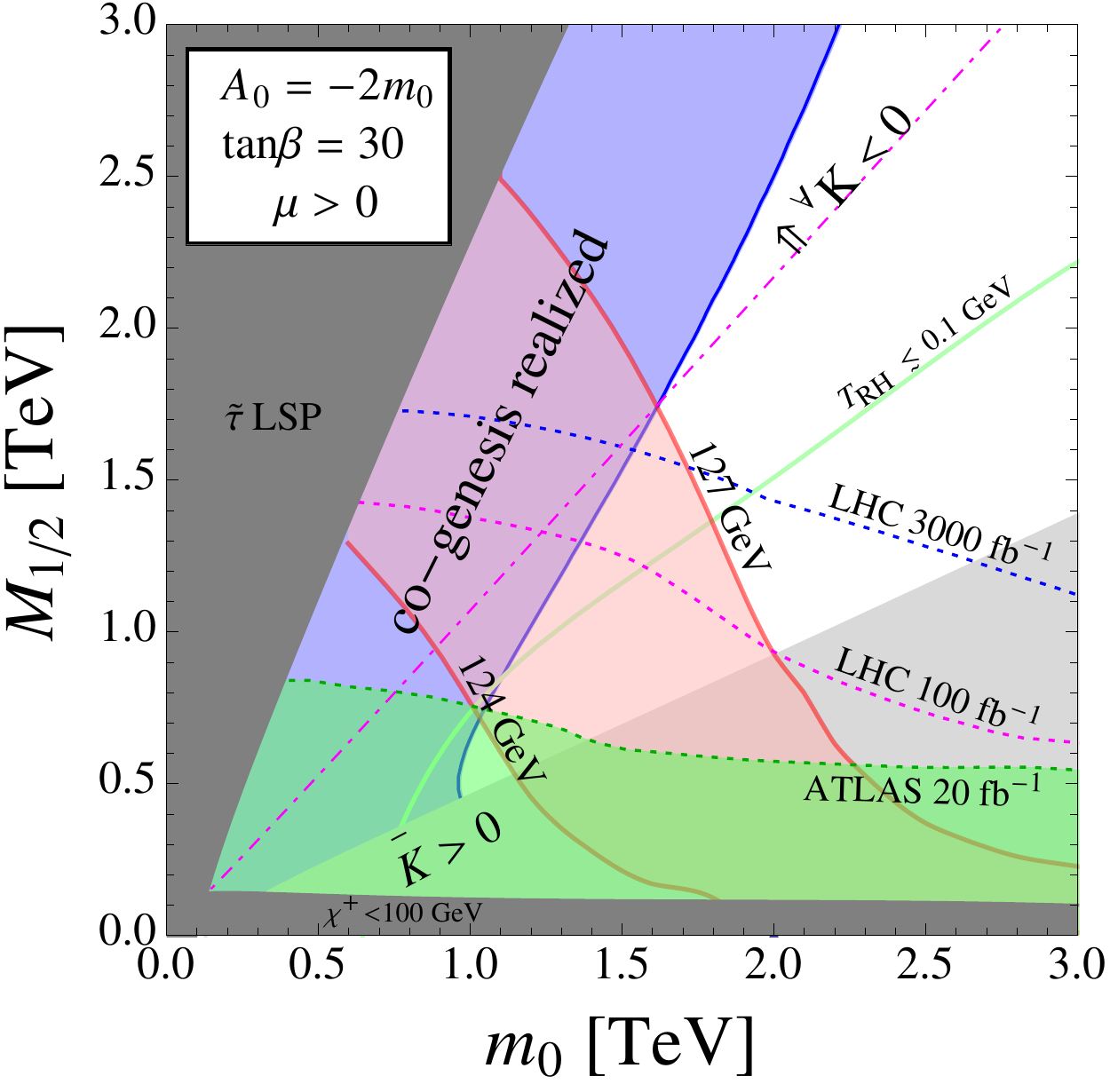} 
\caption{
Allowed contours consistent with observations 
in two ($m_0, M_{1/2}$) planes of the CMSSM, 
with tan$\beta = 10$ and $A_0 = 0$ (upper panel) 
and tan$\beta = 30$ and $A_0 = -2 m_0$ (lower panel) 
with sign$[\mu] = + 1$. 
We can account for the baryon-to-DM ratio as well as baryon density 
in the blue shaded region, where we use $n_q \le 36$. 
The red lines represent contours for the mass of the Higgs boson. 
The annihilation of the LSP is ineffective 
above the light green lines, which is plotted in the case of $\TR \lesssim 0.1 \GeV$. 
Above the magenta dot-dashed line, 
$K < 0$ for all squarks. 
In the light gray region, 
the averaged value of $K$ is positive. 
The dark gray shaded areas are excluded 
either because the LSP is charged, 
there is no consistent electroweak vacuum, 
or the mass of chargino is less than $100 \GeV$. 
The light green regions are excluded 
by the ATLAS search. 
The $14 \TeV$ LHC with $100 \fb^{-1}$ 
and $3000 \fb^{-1}$ would 
probe the parameter space below 
the magenta and blue dotted line, respectively~\cite{Feng:2013tvd, Baer:2009dn, Baer:2012vr}. 
We assume that the top quark pole mass as $m_t^{\rm pole} = 173.3 \GeV$. 
}
  \label{fig1}
\end{figure}

Figure~\ref{fig1} shows that 
the observed baryon-to-DM ratio is realized in the blue shaded region and 
it requires $m_0 \sim M_{1/2}$. 
This is because 
the function $f$ in Eq.~(\ref{correction}) has to be suppressed for bino mass of $O(1)$ TeV, 
that is, $m_{\tilde{b}} \simeq \omega_0$ $(\simeq m_0)$. 
The annihilation of the LSP is ineffective 
above the light green lines, which 
we assume $\TR \lesssim 0.1 \GeV$. 
For larger reheating temperature, 
the annihilation of the LSP is more ineffective. 
One can see that the annihilation effect is irrelevant in most of the blue shaded regions 
and Eq.~(\ref{bDMratio}) is justified. 
Above the magenta dot-dashed line, 
$K < 0$ for all squarks, which means that 
Q-balls are always formed after the Affleck-Dine baryogenesis. 
In the light gray region, 
the averaged value of $K$ is positive and 
Q-balls cannot be formed 
unless the flat direction consists mainly of first and second family squarks. 
The red curves are contours of Higgs mass 
calculated with the code ${\tt FeynHiggs~2.10.0}$~\cite{Heinemeyer:1998yj, Heinemeyer:1998np, 
Degrassi:2002fi, Frank:2006yh, Hahn:2013ria}. 
Note that there are uncertainties for the predicted Higgs mass 
coming mainly from the uncertainties for the top mass 
and higher loop corrections. 
The light green regions are excluded 
by the ATLAS search for $\ensuremath{/ \hspace{-.7em} E_T}$ events 
with $20 \fb^{-1}$ of data at $8 \TeV$~\cite{Buchmueller:2013psa}, 
which has been shown to be independent of tan$\beta$ and $A_0$~\cite{Buchmueller:2012hv}.

We also use the code ${\tt micrOMEGAs~3.6.9}$ to calculate the spin-independent 
interactions of the LSP on nucleons~\cite{Belanger:2013oya}. 
We find that 
in the blue shaded regions
the cross section is much less than $10^{-46} \text{ cm}^2$, 
which is out of reach of the XENON1T experiment~\cite{Aprile:2012zx}. 
This is a unique prediction of our co-genesis scenario from the conventional thermal relic scenario, 
where a sizable higgsino mixing usually leads to a detectable signal in the XENON1T in the bulk 
parameter region.

\vspace{0.2cm}
\noindent {\bf Conclusions. }
We have investigated a scenario for co-genesis of baryon and DM from Q-ball decay. 
Since the branchings into quarks and SUSY particles from Q-ball decay 
are related with each other with a simple relation due to the Pauli exclusion principle, 
the resulting baryon-to-DM ratio naturally results in $O(1)$. 
We have applied the calculation to the CMSSM and 
have identified a parameter region in which 
the baryon-to-DM ratio, their absolute abundance, and the Higgs mass 
are consistent with the observations. 
A part of the parameter region would be tested by future LHC experiments (see Fig.~\ref{fig1}).

\vspace{0.2cm}
%
\noindent {\bf Acknowledgements }
This work is supported by
Grant-in-Aid for Scientific research 
from the Ministry of Education, Science, Sports, and Culture (MEXT), Japan, No. 25400248 (M.K.), No. 21111006 (M.K.);
the World Premier International Research Center Initiative (WPI Initiative), 
MEXT, Japan (A.K., M.K., and M.Y.);
the Program for Leading Graduate Schools, MEXT, Japan (M.Y.);
and JSPS Research Fellowships for Young Scientists (M.Y.).
%



\end{document}